\documentclass[11pt]{article}

\usepackage[english]{babel}
\usepackage{amssymb}
\usepackage{epsfig}
\usepackage{graphicx}
\usepackage{parskip}
\usepackage{ntheorem}
\usepackage{amsmath}
\usepackage{verbatim}
\usepackage{centernot}
\usepackage[T1]{fontenc}

\usepackage[style=ieee]{biblatex}
\newcommand{\BlackBox}{\rule{1.5ex}{1.5ex}}

\newtheorem{example}{Example}
\newtheorem{theorem}{Theorem}

\newtheorem{definition}[theorem]{Definition}

\newcommand{\ci}{\perp\!\!\!\perp}
\newcommand{\nci}{\centernot{\ci}}
\newcommand{\n}{\hfill\break}

\begin{document}

\title{Causal programming: inference with structural causal models as finding instances of a relation}

\author{Joshua Brul\'{e}\footnote{Department of Computer Science,
University of Maryland, College Park. \texttt{jbrule@cs.umd.edu}}}

\date{}

\maketitle

\begin{abstract}
This paper proposes a causal inference relation and causal programming as general frameworks for causal inference with structural causal models. A tuple, $\langle M, I, Q, F \rangle$, is an instance of the relation if a formula, $F$, computes a causal query, $Q$, as a function of known population probabilities, $I$, in every model entailed by a set of model assumptions, $M$. Many problems in causal inference can be viewed as the problem of enumerating instances of the relation that satisfy given criteria. This unifies a number of previously studied problems, including causal effect identification, causal discovery and recovery from selection bias. In addition, the relation supports formalizing new problems in causal inference with structural causal models, such as the problem of research design. Causal programming is proposed as a further generalization of causal inference as the problem of finding optimal instances of the relation, with respect to a cost function. 
\end{abstract}

\section{Introduction}

The development of formal reasoning about causality has roots in philosophy, statistics, economics, and computer science, which has produced a wide variety of different and conflicting terminology. There are many different notations in use for the same concepts, but also many cases of different concepts being referred to by the same name.

While a `grand unified theory' of causation is well beyond the scope of a single paper, this paper suggests that the proposed causal inference relation and causal programming frameworks unify a large number of problems in causal modeling and inference that were previously treated as separate problems. The frameworks also yield several extensions to existing problems, and suggests how all of these problems may be amenable to automated inference.

This paper is part survey and part proposal and only assumes background knowledge of basic probability theory and statistics. The sections ``Desiderata for causal modeling and inference'', ``Potential outcomes'', ``The Heckman hierarchy'', ``Structural causal models'', ``Marschak's maxim and causal diagrams'', and ``The causal hierarchy'' discuss the potential outcome and structural causal model approaches to causality and how they relate to each other. Readers familiar with potential outcomes (Rubin causal model) and structural causal models may wish to begin with ``Causal inference as a logical relation'', which introduces the proposed causal inference relation as a general framework for studying problems in causal inference. ``Restricted causal inference relation'' describes domains for the relation that unify several previously studied problems. ``Other domains for the causal inference relation'' gives examples of how the relation can be extended to represent additional problems of interest. Finally, ``Causal programming'' is proposed as a further generalization which casts causal inference problems as special cases of a general optimization problem.

\section{Desiderata for causal modeling and inference}

Causality is an intuitive idea that can be difficult to rigorously formalize. Causality is implicit in ordinary language \autocite{brown1983}, which makes it easy to accidentally introduce unwarranted assumptions, or fail to introduce necessary assumptions in analysis.

A core concept in causality is the idea that certain variables will respond to changes in other variables. Consider Newton's second law, relating force, mass and acceleration, which simple algebra permits being written in three different ways:

\[ F = m a \]
\[ m = \frac{F}{a} \]
\[ a = \frac{F}{m} \]

Common intuition suggests that if the force applied to some object were increased, it would experience greater acceleration; the mass of the object would not spontaneously increase to compensate. In other words, force causes acceleration, but does not causally effect changes in mass. This is not clear from the standard presentation of the equations, treating the equals sign as the equality relation, where all three equations are equivalent. Under this interpretation of the equals sign, the equations specify a relationship that must be satisfied, but do not specify how the system would respond to an external action.

The three equations are different if the equals sign is treated as an assignment operator --- this interpretation may be more familiar to programmers.\footnote{One might observe that the meaning would be clearer if a different symbol such as \texttt{:=} or \texttt{<-} was used to distinguish assignment from equality, but this convention has not been widely adopted in causal modeling.} In this case, only the third equation captures the intuition that if force were increased, then acceleration would increase proportionally, as long as no other changes to the system were made. Likewise, if mass were increased, acceleration would decrease proportionally. Assumptions of this kind are commonly referred to as \emph{ceteris paribus} assumptions, literally, ``other things being equal''. In particular, \emph{ceteris paribus} is a mainstay of economic analysis \autocite{heckman2005}.

This view of causality has its roots in neoclassical economics, especially in the work of Mill \autocite{mill1848} and Marshall \autocite{marshall1890} and was made more precise with Haavelmo's account of (linear) structural equation models \autocite{haavelmo1943}. Rubin and Holland \autocite{holland1986} provide a pithy motto that summarizes the main idea:

\begin{quote}
    No causation without manipulation
\end{quote}

This account of causality may not seem entirely satisfactory, depending on how `manipulation' is interpreted. For example, few would object to thinking of the Sun's gravity as a cause of Earth's orbit, although, in practice, there are a number of obstacles to significantly manipulating the Sun's mass. This does not change the expectation that if the Sun's mass were suddenly zero, then the Earth would not continue to orbit, no matter how implausible actually implementing such a change would be.

Implicit in such examples of causation is the idea that some effect ($Y$), \emph{could have been different} if a cause ($X$) had been different, regardless of what was actually observed. Such hypotheticals are usually referred to as `counterfactuals'. The idea of defining causality in terms of counterfactuals originates with Hume, defining a cause to be, ``\ldots where, if the first object had not been, the second never had existed'' \autocite{hume1748}. This idea was made more precise with Lewis' account of counterfactuals, using the possible world semantics of modal logic \autocite{lewis1973}:

\begin{quote}
If $c$ and $e$ are two distinct actual events such that $e$ would not have occurred without $c$, then $c$ is a cause of $e$.
\end{quote}

A classic example of a counterfactual sentence is, ``If Nixon had pressed the button, there would have been a nuclear holocaust'' \autocite{fine1975}, which features a number of important characteristics of counterfactuals in general. It cannot (nor should) be empirically tested, but it is still related to the observable world --- note that Nixon did not ever order a nuclear strike, nor is the world a nuclear wasteland. The sentence also, indirectly, implies empirical consequences. If the original counterfactual sentence is true, anyone acting in a sufficiently similar scenario who does `press the button' should expect a nuclear holocaust.

The ideas of manipulations and counterfactuals are related. One view is that a manipulation changes the original system or, more abstractly, generates a new model that represents the effects of the change. Alternatively, the complete set of counterfactuals can be thought of as existing \emph{a priori} and related to observable variables by consistency constraints. Whether or not counterfactuals `actually' exist need not be a concern. Heckman summarizes the relevant metaphysical concerns as, ``A model is in the mind. As a consequence, causality is in the mind'' \autocite{heckman2005}.

Crucially, the notion of counterfactuals is distinct from that of uncertainty --- note that there is no notation in probability theory for ``would have been''. At the same time, statements of causality often include a probabilistic aspect. For example, most would interpret, ``If the grass is wet, then it rained'', as a statement that it is likely that rain caused grass to be wet, without committing to a fully deterministic model such as Newton's second law does.

To summarize, a satisfactory approach to causal modeling and inference requires the distinct concepts of manipulation, counterfactuals and uncertainty; treating any one of these as the same concept is a category error. In addition, such an approach should concur with human intuition as much as possible --- a difficult to understand formalism makes it difficult for subject-matter experts to formalize their domain knowledge.

\section{Potential outcomes (Rubin causal model)}

The potential outcome approach to causal inference (also known as the Rubin causal model, or the Neyman-Rubin-Holland model) provides a notation suitable for representing counterfactual statements.
The statement that ``$Y$ would have taken on value $y$, if $X$ had been $x$, for unit $u$'' is written as:

\[ Y_{x}(u) = y \]

Units are primitives in the potential outcomes approach, which does not define them further. Examples of units include individual patients in a clinical setting, or individual plots of land in an agricultural study. Variables (e.g. $X$, $Y$) are real-valued functions defined for every unit; for example, $X$ is commonly defined to be treatment and $Y$ defined to be response to treatment in a clinical setting.

A particular quantity of interest is treatment effect\footnote{This quantity is sometimes referred to as `causal effect' \autocite{holland1986}; this paper adopts the convention of referring to this as `treatment effect' to avoid confusion with the structural causal model definition of causal effect.}, which is defined as the difference in response when a particular unit is exposed to treatment ($X=t$) versus control ($X=c$):

\[ Y_t(u) - Y_c(u) \]

Causal inference is difficult because, although there are many potential outcomes for any particular variable, it is only possible to observe one \emph{actual} outcome. For example, it is impossible to treat and not treat the same patient.\footnote{One might object that it is possible to not treat a patient initially, and then treat the same patient later, but these are different units. The patient's condition after waiting long enough to observe the effects of non-treatment is different than their initial condition.} Holland summarizes this as the Fundamental Problem of Causal Inference:

\begin{quote}
    It is impossible to \emph{observe} the value of $Y_t(u)$ and $Y_c(u)$ on the same unit and, therefore, it is impossible to \emph{observe} the effect of $t$ on $u$.
\end{quote}

Causal inference is impossible without making additional assumptions --- data alone provides no knowledge of how observations will generalize to other circumstances. An example of a simple assumption that makes causal inference possible is unit homogeneity, which can be thought of as `laboratory conditions'. If different units are carefully prepared, it may be reasonable to assume that they are equivalent in all relevant aspects, i.e. $Y_t(u_1) = Y_t(u_2)$ and $Y_c(u_1) = Y_c(u_2)$. For example, it is often assumed that any two samples of a given chemical element are effectively identical. In these cases, treatment effect can be calculated directly as $Y_t(u_1) - Y_c(u_2)$. However, it is often the case that such tightly controlled conditions are impossible to maintain. Accordingly, the main focus of the potential outcomes approach is on average effects.

A probability distribution over the universe of units, $P(u)$, induces a probability distribution over the potential outcome variables. Formally \autocite{pearl2009}:

\[ P(Y_{x} = y) = \sum_{\{u \:\mid\: Y_{x}(u)=y\}} P(u) \]

Since the potential outcome variables are random variables, it is meaningful to speak of average treatment effects. In particular, expected value ($E$) is a linear operator which permits writing:

\[ E( Y_t - Y_c ) = E(Y_t) - E(Y_c) \]

In other words, it is possible to estimate average treatment effect by estimating $E(Y_t)$ and $E(Y_c)$ individually. Unfortunately, it is not possible to sample from either of these random variables directly. $Y_t$ is treatment over the \emph{entire} universe of units, a counterfactual world where every patient was exposed to treatment. Actual samples would be from the random variable $Y$. Although these variables are different, they are still related to each other by the consistency rule \autocite{galles1998, gibbard1978-counterfactuals}:

\[ X(u) = x \implies Y(u) = Y_x(u) \]

In other words, if the variable $X$ is observed to take on value $x$, then the potential response $Y_x$ is simply the current value of $Y$. An immediate consequence of consistency is: $P(Y_x = y \mid X=x) = P(Y=y \mid X=x)$. Furthermore, if response to treatment $Y_x$ is independent of treatment \mbox{($Y_x \ci X$)}, then the following equalities hold:

\[ E(Y \mid X=t) = E(Y_t) \]
\[ E(Y \mid X=c) = E(Y_c) \]

In this case, average treatment effect can be estimated directly from the collected samples as $E(Y \mid X=t) - E(Y \mid X=c)$. This expression is sometimes referred to as the \emph{prima facie} treatment effect \autocite{holland1986}.

However, there are many scenarios where selection of treatment is not independent of response to treatment. Consider the question of whether smoking is a cause of cancer. The \emph{prima facie} effect may be significant. However, it is conceivable that there exists a latent genetic factor that predisposes individuals to smoke, and also makes them more susceptible to cancer.\footnote{Statistician Ronald Fisher is infamous for having spoke out against studies linking smoking to cancer, while being ardent tobacco user himself. He later died of cancer. However, his actual objections to the studies were not incorrect.} This is an example of the well-known problem of confounding variables and the possibility of latent confounding variables is especially difficult to rule out.

The potential outcomes approach generally operates by making independence assumptions about potential outcome variables. Randomization, i.e. the samples were obtained in a randomized controlled trial, makes the assumption\footnote{This condition is referred as `no confounding', `exogenity' or `ignorability', depending on the source \autocite{pearl2009, rosenbaum1983, engle1983-exogeneity}.} $Y_x \ci X$ especially plausible since --- at least, theoretically --- the selection of treatment or non-treatment for each unit is determined entirely by an independent source of randomness. In practice, there may be issues with imperfect compliance (i.e. some patients may fail to take the drugs they are assigned), but randomized controlled trials remain the `gold standard' for causal evidence.

Another common type of assumption that can permit causal inference is conditional ignorability, $(Y_x \ci X \mid Z)$, which is the statement that $Y_x$ and $X$ are conditionally independent given $Z$, a set of covariates that are being `adjusted' or `controlled' for. For example, if it were known that there was a genetic factor $Z$ (and no \emph{other} such factors) that caused both smoking and cancer, then it would be reasonable to assume conditional ignorability, which would permit the following derivation \autocite{pearl2009}:

\[
    \begin{split}
        P(Y_x = y) & = \sum_z P(Y_x = y \mid z) P(z)\\
                   & = \sum_z P(Y_x = y \mid x, z) P(z)\\
                   & = \sum_z P(Y = y \mid x, z) P(z)\\
                   & = \sum_z P(y \mid x, z) P(z)
    \end{split}
\]

The notation belies a fundamental shift in perspective: the formula computes a \emph{potential outcome} variable, $P(Y_x)$, entirely in terms of \emph{observable} probabilities, $P(y \mid x, z)$ and $P(z)$, with respect to the assumption of conditional ignorability.

There are two problems that have not been directly addressed. First, in practice, an analyst only has access to a finite-sample distribution $\hat{P}$, instead of the true population distribution $P$. The other problem lies in justifying the conditional independence assumptions used in analysis. For example, what does it mean for conditional ignorability to be true, and how would an analyst judge if such an assumption is reasonable?

\section{The Heckman hierarchy}

Reasoning about causality benefits from maintaining clear distinctions between different tasks in causal inference. In particular, Heckman \autocite{heckman2005} identifies three distinct tasks in causal inference that are often conflated in practice:

\begin{enumerate}
    \item
        Definitions of counterfactuals

    \item
        Identification of models/parameters from population distributions

    \item
        Selection of models from real data (i.e. sampling distributions)

\end{enumerate}

The potential outcomes approach, combined with other statistical tools, provides a way of performing all of these tasks. The potential outcome syntax permits analysts to formally write independence assumptions about potential outcome variables, implicitly defining the nature of the counterfactuals. The rules of probability theory combined with algebraic manipulations (may) permit identification of quantities of interest from population distributions, e.g. $P(Y_x = y) = \sum_z P(y \mid x, z) P(z)$, assuming conditional ignorability. Finally, estimation and hypothesis testing theory provide a way to calculate confidence intervals from sampled data and accept or reject hypotheses appropriately.

However, the potential outcomes notation, alone, provides an incomplete definition of counterfactuals. It provides a formal way of writing, for example, that response to treatment ($Y_x$) is independent of treatment ($X$), but does not provide a formal definition of what it means for this assumption to be true or false. In the language of formal logic, potential outcomes notation provides \emph{syntax} but not \emph{semantics} for causal statements. Giving these statements meaning requires formalizing the notion of a causal model.

\section{Structural causal models}

Consider a simple economic model of propensity to consume, assuming all prices are constant. As an example, Haavelmo suggests a model where, ``if the group of all consumers in society were repeatedly furnished with the total income or purchasing power $x$ per year, they would, on the average or `normally,' spend a total amount $y$ equal to'' \autocite{haavelmo1943}:

\[ y = \beta x + \alpha \]

Where $\alpha$ and $\beta$ are constants. Naturally, it would be unreasonable to expect that, in any particular year, spending would be exactly equal to $y$. This is not merely a consequence measurement errors --- presumably there are a large number of additional factors that could affect spending that are not directly accounted for in this simple model. These additional factors can be indirectly represented by adding a residual or `error term' to the original equation:

\[ y = \beta x + \alpha + \epsilon \]

Where $\epsilon$ is random variable with mean value zero, regardless of the value of $x$. This is a simple example of a structural equation model (SEM).

Haavelmo is notable for being among the first to explicitly interpret such equations as predicting the result of idealized experiments. It may be the case that an analyst is merely trying to fit the equation to the past and hopes that the relation holds in the future, assuming no significant changes to the underlying system. A stronger assumption is that consumers will continue to respond in the same way to income, regardless of the sources from which their income originates. With respect to this assumption, it is possible to predict the result of an intervention (e.g. government spending or taxation) to set income at a given level. Pearl further formalizes this interpretation of structural equations:\n

\begin{definition}[Structural Equations \autocite{pearl2009}]
    An equation $y = \beta x + \epsilon$ is said to be \emph{structural} if it is to be interpreted as follows: In an ideal experiment where we control $X$ to $x$ and any other set $Z$ of variables (not containing $X$ or $Y$) to $z$, the value $y$ of $Y$ is given by $\beta x + \epsilon$ where $\epsilon$ is not a function of the settings $x$ and $z$.
\end{definition}

Note that this definition assumes an \emph{ideal} experiment to control $X$. Many manipulations that are theoretically simple can turn out to be difficult or impossible to implement in practice. This is not a strike against the definition, but a warning to carefully model interventions as well as the causal relationships themselves.

The philosophy of causality adopted here is that of Laplacian (quasi-) determinism. The residual, $\epsilon$, represents all of the additional factors that determine $Y$ that are not directly modeled. In principle, if these factors were completely known, it would be possible to exactly determine how $Y$ would respond to any change. In this view, randomness is a statement of analyst's ignorance, not inherent to the system itself.

This is related to the potential outcomes approach, which considers potential outcome variables to be real-valued functions of `units'. Since units are primitives and not defined further, these functions are implicit. In comparison, structural equation modeling works with explicit functions, where variables are functions of all of their determining factors.

One of the weaknesses of structural equation modeling is that it makes very strong assumptions --- usually, linearity and the assumption that all variables are multivariate normal. It is perhaps unsurprising then that many analysts are reluctant to assign causal meaning to the equations and consider them to be merely a `shorthand' way to represent a joint probability distribution. The linearity assumption, in particular, is very restrictive --- the earlier example of Newton's second law violates it. Consider, also, the smoking/cancer example, where $X$ is smoking, $Y$ is cancer, and $Z$ is a possible genetic factor that predisposes one to smoke and can cause cancer. These assumptions can be captured in the following three equations:\n

\begin{example}[Smoking/cancer model]
\[ Z = f_Z(\epsilon_Z) \]
\[ X = f_X(Z, \epsilon_X) \]
\[ Y = f_Y(X, Y, \epsilon_Y) \]
\end{example}

Where each $f_i$ is some --- likely nonlinear -- function. $X, Y, Z$ are called `endogenous variables' since they are determined by factors in the model. $\epsilon_X, \epsilon_Y, \epsilon_Z$ are called `background variables' since they are determined by outside factors that are not directly accounted for.\footnote{These are sometimes referred to as `exogenous' variables. Unfortunately, `exogeneity' is often used to refer to a number of subtly different conditions between sets of variables in a causal model. To avoid confusion, the term `background variable' will be used in this paper.}

This nonlinear generalization of structural equation models with arbitrarily distributed background variables originates with Pearl and Verma \autocite{pearl1991} and has been referred to by several different names including `probabilistic causal models', `graphical causal models' and     `structural causal models'. The name `structural causal models' will be used in this paper, since it appears to be least likely to name clash with other terms in the literature.\n

\begin{definition}[Structural Causal Model \autocite{bareinboim2014-thesis}]

    A \emph{structural causal model} M is a tuple $M = \langle U, V, F, P(u) \rangle$, where:

    \begin{enumerate}
        \item
            A set $U$ of background (also called \emph{exogenous}) variables, that are determined by factors outside the model

        \item
            A set $V = \{ V_1, \ldots V_n \}$ of variables, called \emph{endogenous}, that are determined by variables in the model --- that is, variables in $U \cup V$;

        \item
            $F$ is a set of functions $\{ f_1, \ldots, f_n \}$ such that each $f_i$ is a mapping from (the respective domains of) $U_i \cup PA_i$ to $V_i$, where $U_i \subseteq U$ and $PA_i \subseteq V \setminus V_i$ and the entire set $F$ forms a mapping from $U$ to $V$;

        \item
            $P(u)$ is a probability function defined over the domain of $U$. \n

    \end{enumerate}

\end{definition}

Note that the definition of structural causal models requires that the set of equations, $F$, form a mapping from $U$ to $V$. In other words, $F$ has a unique solution for $V$ as a function of $U$. A sufficient condition for this is that the system is recursive, i.e. there are no cyclic dependencies in the parent ($PA_i$) sets of the endogenous variables. A key difficulty with nonrecursive systems in structural causal models is that they may require solving systems of nonlinear equations; this paper will not consider nonrecursive systems.

Structural causal models provide a straightforward definition of interventions. Consider an action to force some set of variables $X$ to take on particular values $x$; this is represented using the $do()$ operator. \n

\begin{definition}[Effect of action \autocite{pearl2009}]

    Let $M$ be a causal model, $X$ a set of variables in $V$, and $x$ a particular realization of $X$. The \emph{effect of action} $do(X=x)$ on $M$ is given by the submodel $M_x$.\n

\end{definition}

\begin{definition}[Submodel \autocite{pearl2009}]

    Let $M$ be a causal model, $X$ a set of variables in $V$, and $x$ a particular realization of $X$. A submodel $M_x$ of $M$ is the causal model:

    \[ M_x = \langle U, V, F_x, P(u) \rangle \]

    Where:

    \[ F_x = \{ f_i : v_i \notin X \} \cup \{ X=x \} \]

\end{definition}

A submodel produced by $do(X=x)$ can be thought of as the result of `wiping out' each $f_i$ that determines each $X_i$, and replacing $f_i$ with the constant $x_i$, a process which Pearl colorfully refers to as performing ``surgery on equations'' \autocite{pearl2009}. As an example, consider an idealized intervention to determine the causal effect of smoking on cancer. In the original model, $M$, the decision to smoke ($X$) is a function of a background variable ($\epsilon_X$) and a genetic factor ($Z$) that both predisposes one to smoke and affects cancer risk. The intervention, $do(X=x)$, effectively `cuts out' the confounding from the genetic factor and produces a new model, $M_x$, in which the factors that determine $Z$ and $Y$ are unchanged, but $X$ has been set to the value $x$: \n

\begin{samepage}
\begin{example}[Smoking/cancer submodel]
\[ Z = f_Z(\epsilon_Z) \]
\[ X = x \]
\[ Y = f_Y(X, Y, \epsilon_Y) \]
\end{example}
\end{samepage}

Given the definitions of a submodel and effect of action, the relationship between potential outcomes and structural causal models is remarkably straightforward:\n

\begin{definition}[Potential Response \autocite{pearl2009}]
    Let $X$ and $Y$ be two subsets of variables in $V$. The \emph{potential response} of $y$ to action $do(X=x)$, denoted $Y_x(u)$, is the solution for $Y$ of the set of equations $F_x$, that is, $Y_x(u) = Y_{M_x}(u)$.\n
\end{definition}

The probability of $y$, given the action $do(X=x)$ is denoted\footnote{Other notations, including $P_x(y)$ and $P(y \mid \hat{x})$ are in use, but will not be used in this paper. In particular, the `hat' notation risks confusion with the standard statistical practice of denoting estimates of random variables with a hat.} by either $P(y \mid do(x))$ or $P(Y_x)$ and is induced by the probability distribution over the background variables, $P(u)$, and the submodel, $M_x$:

\[ P(Y_x = y) = P(y \mid do(x)) = \sum_{ \{ u \:\mid\: Y_{M_x}(u) = y \} } P(u) \]

This establishes the theoretical connection between potential outcomes and structural causal models. It also highlights the philosophical differences between traditional structural equation modeling and potential outcome analysis. In structural equation modeling, equations are usually assumed to be linear, with the random variables being multivariate normal. In other words, structural equation modeling relies on strong and explicit model assumptions. Potential outcome analysis is effectively the opposite; the model assumptions are weak and implicit. Independence assumptions between potential outcome variables implicitly constrain the set of possible causal models under consideration, but do not provide much guidance on determining what that set is.

\section{Marschak's maxim and causal diagrams}

Heckman coined `Marschak's Maxim', in honor of an insight by Marschak \autocite{marschak1953}:

\begin{quote}
Forecasting policies may require only partial knowledge of the system.
\end{quote}

From the definition, a complete specification of a structural causal model requires specifying the functions that determine each endogenous variable and the probability distribution over the background variables. The former is often difficult to know; the later is often impossible, considering that the background variables are usually the very factors that cannot be directly accounted for.

Marschak's maxim is a reminder that a partial specification of a model may still be sufficient to conduct causal inference. A set of independence assumptions made in a potential outcomes analysis implicitly denotes a set of possible models. Causal diagrams are another approach.

\begin{figure}[h]
    \centering
    \includegraphics{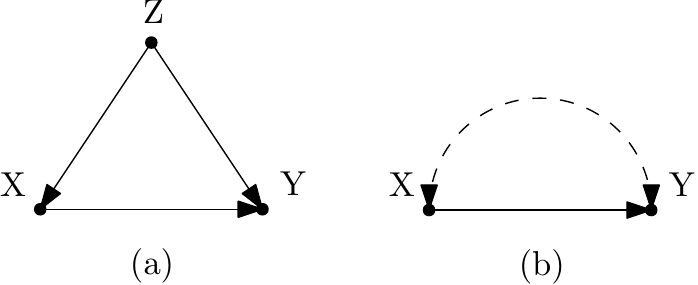}

    \caption{Markovian and semi-Markovian causal diagrams}
    \label{fig:markovian}
\end{figure}

Every causal model induces a causal diagram, where each vertex in the diagram corresponds to an endogenous variable, $V_i$, and directed edges point from members of $PA_i$ to $V_i$. If the background variables are jointly independent and each background variable appears in only one $PA_i$ set, the model is called Markovian and the joint probability function, $P(v)$, will respect the Markov condition, i.e. every variable is conditionally independent of its non-descendants, given its parents \autocite{pearl2009}. Otherwise, the model is called semi-Markovian. Dependencies between endogenous variables due to background variables are denoted by dashed, bidirectional edges.\footnote{An alternative convention is to enter observable variables as solid nodes and latent variables as hollow nodes.} For example, if the genetic factor in the smoking/cancer example were known and measurable, the model would be Markovian; otherwise, the model would be semi-Markovian and $X$ and $Y$ would have a dashed, bidirectional edge between them to denote the dependency (figure \ref{fig:markovian}).

There is a useful correspondence between causal diagrams and causal models. Every causal model induces a causal diagram, and every causal diagram has at least one model (in fact, infinitely many) that would induce it. In this sense, a casual diagram can be thought of as denoting a set of models where each endogenous variable is assumed to be a function its parents, without committing to an assumption of what the function is.

Inference in the structural causal model approach is generally performed with respect to the assumptions entailed by a causal diagram. For example, calculating the causal effect of $X$ on $Y$ in the smoking/cancer model is a simple adjustment for direct causes.\n

\begin{theorem}[Adjustment for Direct Causes \autocite{pearl2009}]

    Let $PA_i$ denote the set of direct causes of variable $X_i$ and let $Y$ be any set of variables disjoint of $\{ X_i \cup PA_i \}$. The effect of the intervention $do(X_i = x_i)$ on $Y$ is given by:

    \[ P(Y \mid do(x_i)) = \sum_{pa_i} P(y \mid x_i, pa_i) P(pa_i) \]

\end{theorem}

Applying this theorem to the smoking/cancer example with observable $Z$ yields:

\[ P(y \mid do(x)) = \sum_z P(y \mid x, z) P(z) \]

Formally, it is said that the probabilities $P(y \mid x, z)$ and $P(z)$ and the causal diagram identify $P(y \mid do(x))$. Note that if $Z$ is latent, then $P(y \mid do(x))$ is not identifiable. Intuitively, there is no way of knowing if correlation between $X$ and $Y$ is due to the latent factor, or due to the effect of $X$ on $Y$.

Unsurprisingly, this is the same result as from the potential outcomes analysis. In the potential outcomes approach, the set of causal models under consideration is implicitly specified by the conditional independence assumptions between potential outcome variables. Causal diagrams more explicitly denote the set of models under consideration, and consider properties like conditional independences to be a consequence of the model assumptions entailed in the diagram. In both cases, Marschak's maxim is in play. A complete specification of the model is not needed to calculate the causal effect; the formula correctly calculates $P(y \mid do(x))$ for all models under consideration.

\begin{figure}[h]
    \centering
    \includegraphics{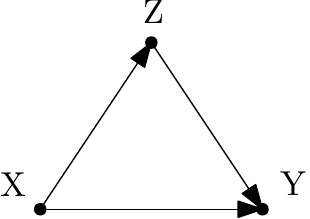}

    \caption{Blood pressure causal diagram}
    \label{fig:mediated}
\end{figure}

Note that incorrectly adjusting for variables can produce biased estimates of causal effect. Consider a model where $X$ is treatment, $Y$ is recovery, and $Z$ is blood pressure, which causally affects recovery, and is affected by treatment (figure \ref{fig:mediated}). An adjustment for direct causes should not be performed in this case, since $Z$ is not a direct cause of $X$. Intuitively, adjusting for blood pressure `blocks' the causal effect of $X$ on $Y$ that is mediated through $Z$.
 
\section{The causal hierarchy}

The previous sections have focused on examples of identifying causal effects, e.g. ``If a patient is treated, will they recover?'' Queries about causal effects are only one of the possible queries in the full causal hierarchy \cite{shpitser2008}:

\begin{enumerate}

    \item
        Statistical/associational: queries involving no interventions. For example, ``If I \emph{observe} that a patient has been exposed to $X=x$, how likely is it that they will recover ($Y=y$)?'' i.e. $P(y \mid x)$

    \item
        Causal/interventional: queries about the result of outside interventions. For example, ``If I \emph{treat} a patient, how likely is it that they will recover?'' i.e. $P(y \mid do(x))$ or $P(Y_x = y)$

    \item
        Counterfactual: queries involving multiple hypothetical worlds. For example, ``Given that a patient was exposed to $x$, how likely is it that they \emph{would have} recovered, even if they had not been exposed ($X=x'$)?'' i.e. $P(Y_{x'} = y \mid x)$

\end{enumerate}

This forms a hierarchy, in that each successive class of queries includes the previous as a special case. Statistical queries only involve the original model; a joint probability distribution is sufficient to compute such a query. Causal queries are questions about a system after an external intervention, i.e. queries about the submodel resulting from a $do()$ action. The combination of the pre-intervention joint probability distribution and a Markovian causal diagram is sufficient to answer causal queries \autocite{pearl2009}. Finally, counterfactual queries span multiple models. A counterfactual query can contain `conflicting' information, asking about the same variable taking on different values in different hypothetical worlds. For example, $P(Y_{x'}=y \mid x)$ is called the effect of treatment on the treated \autocite{shpitser2009} and considers two `parallel worlds': one where the patient was treated and one where the patient was not. Although these are different models, they share the same background variables and functional relationships (i.e. the $f_i$ equations). Intuitively, the mechanisms that determine whether or not a patient recovers remain the same across the hypothetical worlds, even if the assignment of treatment/non-treatment does not.

In general, answering counterfactual queries requires knowledge of the functional relationships and information about the distribution of the background variables. Note that even data from a randomized controlled trial may be insufficient, since it is impossible to simultaneously treat and not treat the same patient. However, there are special cases in which counterfactual queries can be identified from less complete knowledge. For example, there exists a graphical criterion for identifying the effect of treatment on the treated \autocite{shpitser2009}. With respect to certain causal diagrams, it is possible to determine the effect of treatment on the treated from the pre-intervention distribution and the diagram's causal assumptions alone.

\begin{figure}[h]
    \centering
    \includegraphics{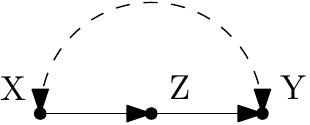}

    \caption{A causal diagram where $P(Y_x = y \mid x') = \sum_z P(y \mid z, x) P(z \mid x')$ \autocite{shpitser2009}}
    \label{fig:ett}
\end{figure}

\section{Causal inference as a logical relation}

The main original contribution of this paper is to introduce the causal inference relation:

\[ \langle M, I, Q, F \rangle_V \]

Where:

\begin{itemize}

    \item
        $M$ is a set of \emph{model} assumptions, entailing a set of structural causal models

    \item
        $I$ is a set of \emph{information}; specifically, a set of known population probabilities

    \item
        $Q$ is a \emph{query} from the causal hierarchy

    \item
        $F$ is a \emph{formula} that computes $Q$ as a function of the  information set $I$, that holds true for every model entailed by $M$

    \item
        $V$ is the set of endogenous variables under consideration

\end{itemize}

The causal inference relation is indexed by $V$; there is a relation for each set of endogenous variables. $V$ can be thought of as the set of all variables that can be potentially manipulated and/or measured.

$M$ is a finite set of model assumptions, which may permit denoting an infinite set of structural causal models under consideration. A causal diagram, typically denoted $G$, is a particularly compact representation of model assumptions. In principle, $M$ can be any restriction on the set of all recursive structural causal models; this includes such assumptions as conditional independence assumptions between potential outcome variables, or restrictions on the functional relationships, such as the assumption that all $f_i$ are linear. However, the set of all causal diagrams over $V$ will serve as the main example of the domain of $M$.

The information set, $I$, can be thought of as the data collected about a particular system, albeit idealized, large-sample data, since $I$ is defined to be population, not sample probabilities. A familiar information set is the pre-intervention joint probability function, often referred the `observational' probability distribution. This is typically denoted $P(v)$ or $P$. $I$ may also be extended to include other population probabilities. For example, it may be the case that, in addition to observational data, experimental data for a limited subset of variables is available. Such an $I$ could be represented as $P(v \mid do(z')), \forall Z' \subseteq Z$, where $Z$ is some subset of $V$ that an analyst can directly manipulate and observe the effects of. Note that this includes the observational probability distribution, as $P(v \mid do(z')) = P(v)$ when $Z'$ is the empty set. Unless otherwise stated, it will be assumed that all probability distributions are (strictly) positive, i.e. $P(v) > 0, \forall v$, as this is required by many theorems in causal inference.

A query, $Q$, can be any query from the causal hierarchy. This paper focuses mainly on causal effect queries, e.g. $P(y \mid do(x))$.

Finally, the formula, $F$, computes $Q$ as a function of $I$, in all models entailed by $M$. Note that if the information set is symbolic, then the resulting formula will also be symbolic. If numerical probabilities are available, evaluating the formula will yield the appropriate value for the query. In principle, $F$ could be extended to include bounds on a query, but this paper focuses on exact results.

The simplest problem involving the causal inference relation is determining whether a given tuple $\langle M, I, Q, F \rangle$ is an instance of the causal inference relation. For example, $M = (\text{figure \ref{fig:markovian}a})$, $I=P(x, y, z)$, $Q=P(y \mid do(x))$, and $F= \sum_z P(y \mid x, z) P(z)$ is a valid instance of the relation. The same tuple, but with $M=(\text{figure \ref{fig:mediated}})$ instead is \emph{not} an instance the relation, since $F$ does not correctly compute the query in all models entailed by $M$.

The causal inference relation provides a general framework for studying problems in causal inference. Many problems in causal inference can be seen as finding an instance, or enumerating all the instances, of the causal inference relation that satisfy given criteria. These problems can be broadly categorized by which of $M, I, Q$ are given:

\begin{itemize}

    \item
        $M, I, Q$ - Identification: the problem of finding a formula to compute a causal query

    \item
        $I, Q$ - Causal discovery: the problem of enumerating the models that are compatible with given population probabilities

    \item
        $M, Q$ - Research design: the problem determining the observational and/or experimental information that must be collected to answer a given query

    \item
        $M, I$ - Query generation: the problem of enumerating identifiable queries

\end{itemize}

Note that problems where $F$ is given are not considered, as they represent methodologically suspect practices. For example, searching for $M$, given $I$, $Q$ and $F$ is an attempt to find a \emph{post hoc} rationalization for a calculation of a causal effect that has already been performed.

\subsection{Identification}

Consider a variant of the smoking/cancer model where the effect of smoking ($X$) on lung cancer ($Y$) is mediated through tar deposits in a person's lungs ($Z$) \autocite{pearl1995}. In addition, there may exist a latent factor that directly causes both $X$ and $Y$, but not $Z$. These model assumptions are encoded in the causal diagram in figure 3. Furthermore, suppose that an analyst knows the joint pre-intervention distribution, $P(x, y, z)$, and wishes to compute the causal effect of $X$ on $Y$, $P(y \mid do(x))$.

This corresponds to the following problem: find one instance of the causal inference relation such that $M=(\text{figure \ref{fig:ett}}), I=P(x, y, z), Q=P(y \mid do(x))$. Since $Z$ satisfies the front-door criterion \autocite{pearl1995}, $Q$ can be computed using a front-door adjustment. A full solution to this problem is $\langle M, I, Q, F \rangle$ where $M, I$ and $Q$ are as given, and $F$ is:

\[ \sum_z P(z \mid x) \sum_{x'} P(y \mid x', z) P(x') \]

\begin{figure}[h]
    \centering
    \includegraphics{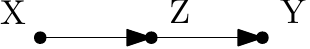}

    \caption{A causal diagram without latent confounding variables}
    \label{fig:unconfounded}
\end{figure}

It is possible for there to be several instances of the causal inference relation that satisfy given criteria. For example, the instances of the causal inference relation that satisfy $M=(\text{figure \ref{fig:unconfounded}})$, $I = P(x, y, z)$, $Q=P(y \mid do(x))$ includes solutions $\langle M, I, Q, F_1 \rangle$ and $\langle M, I, Q, F_2 \rangle$, where $F_1$ is as in the previous example, and $F_2$ is simply $P(y \mid x)$. In an identification problem, an analyst is usually only interested in finding one solution. However, there are other causal inference problems where finding multiple solutions is of interest.

\begin{figure}[h]
    \centering
    \includegraphics{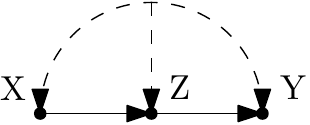}

    \caption{A causal diagram where $P(y \mid do(x))$ is not identifiable}
    \label{fig:fully-confounded}
\end{figure}

Conversely, there may be no instances of the causal inference relation that satisfy given criteria. Searching for instances of the causal inference relation that satisfy $M=(\text{figure \ref{fig:fully-confounded}})$, $I = P(x, y, z)$, $Q=P(y \mid do(x))$ will fail, since $Q$ cannot be uniquely computed in all models entailed by $M$.

Treating the full tuple $\langle M, I, Q, F \rangle$ as the solution --- as opposed to just the formula, $F$ --- may seem redundant for identification problems. The utility of this approach becomes more apparent for less restrictive search criteria.

\subsection{Causal discovery}

If a causal diagram is not specified, then causal inference becomes a problem of causal discovery. As a simple example, consider an analyst that is studying a system with two endogenous variables. Suppose the analyst knows that the variables are dependent, knows the joint observational probability function, i.e. $I=P(x, y)$, where $X \nci Y$, and wishes to infer the causal effect of $X$ on $Y$, i.e. $Q=P(y \mid do(x))$.

\begin{figure}[h]
    \centering
    \includegraphics{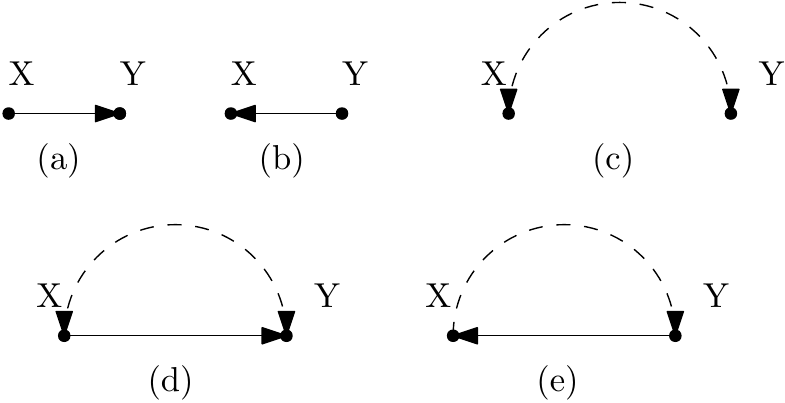}

    \caption{Causal diagrams that are Markov compatible with $I$}
    \label{fig:causal-discovery}
\end{figure}

A causal diagram and probability function are said to be Markov compatible if the probability function respects the conditional independences implied by the Markov condition. There are two causal diagrams that are Markov compatible with $I$ that also permit identification of $Q$: $M_1 = (\text{figure \ref{fig:causal-discovery}a})$ and $M_2 = (\text{figure \ref{fig:causal-discovery}b})$. This corresponds to the following instances of the causal relation: $\langle M_1, I, Q, F_1 \rangle$ and $\langle M_2, I, Q, F_2 \rangle$, where:

\[ F_1 = P(y \mid x) \]
\[ F_2 = P(y) \]

If the domain of $M$ is limited to the space of Markovian causal diagrams, then this set of solutions is also complete, in the sense that every causal diagram that is Markov compatible with $I$ is contained in one of the enumerated instances of causal inference relation. However, if the domain of $M$ also includes semi-Markovian causal diagrams, then there are several causal diagrams that are compatible with $I$ that do not permit identification of $Q$.

Note that any causal diagram where all endogenous variables share a common, latent cause is Markov compatible with every joint observational probability function $P(v)$. This has consequences for interpreting the results of causal discovery. It is generally incorrect to treat causal discovery as definitively determining the causes of variables in a system. Instead, a discovered model can be viewed as a set of additional, compatible assumptions that will permit answering a given query. Causal discovery will usually be incomplete, since non-identifiable models remain a possibility, unless explicitly ruled out by domain knowledge.

Causal discovery usually relies the assumption that $P$ is faithful to $G$ (this condition is also called `stability' \autocite{pearl2009}), which is the assumption that every conditional independence relationship that is true in $P$ is entailed by the Markov condition \autocite{spirtes2000}. For example, if $I=P(x,y)$, where $X \ci Y$, then $P$ is Markov compatible with every diagram in figure \ref{fig:causal-discovery}. However, $P$ is not faithful to any of these diagrams; intuitively, the edges between $X$ and $Y$ suggest a dependency between the variables that is not present.

\subsection{Research design}

\begin{figure}[h]
    \centering
    \includegraphics{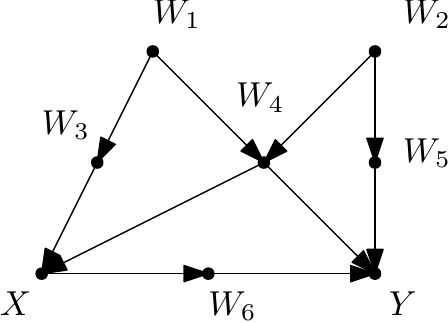}

    \caption{A causal diagram adapted from \autocite{pearl1995}}
    \label{fig:research-design}
\end{figure}

If the information set is not specified, then causal inference becomes a problem of research design. As an example, consider a scenario where an analyst wishes to calculate $P(y \mid do(x))$ with respect to the causal diagram in figure \ref{fig:research-design}.

The complete joint observational probability function $P(v)$ is sufficient, but unnecessary. In particular, an analyst may be interested in calculating causal effect from less information when it is expensive, or otherwise difficult to obtain the complete joint observational probability function. By the back-door criterion \autocite{pearl1995}, $P(y \mid do(x))$ can be computed as either:

\begin{samepage}
\[ F_1 = \sum_{x_3,x_4} P(y \mid w_3, w_4, x) P(w_3, w_4) \]
\[ F_2 = \sum_{x_4,x_5} P(y \mid w_4, w_5, x) P(w_4, w_5) \]
\end{samepage}

Solutions are sensitive to the domain and representation of $I$. One possible representation of $I_1$ is $P(y \mid w_3, w_4, x), P(w_3, w_4)$. However, this implies a somewhat cumbersome domain for $I$ and can make it difficult to determine equivalent information sets. For example, the information set $P(y \mid x), P(x)$ is semantically, but not syntactically, equivalent to $P(x, y)$. A less expressive, but simpler domain for $I$ is the set of joint observational probability functions over subsets of $V$. This domain has a natural partial order: an information set, $P(v_1)$, is included in a more general information set, $P(v_2)$, if $V_1 \subset V_2$. In this context, \emph{minimal} information sets to calculate $P(y \mid do(x))$ are $I_1 = P(x, y, w_3, w_4)$ and $I_2 = P(x, y, w_4, w_5)$.

\subsection{Query generation}

If the query is not specified, then causal inference becomes a problem of query generation. Note that the number of identifiable queries has the potential to be very large. For example, if $I=P(v)$, and $M$ is a Markovian causal diagram, then all queries of the form $P(y_1, \ldots, y_m \mid do(x_1, \ldots, x_n ))$ are identifiable, which is exponential in $\vert V \vert$. Tractable query generation generally requires some restriction on the space of queries or a willingness to accept an incomplete set of solutions.

\begin{figure}[h]
    \centering
    \includegraphics{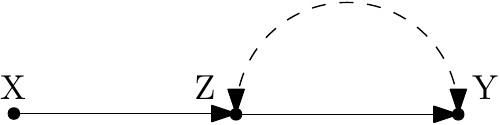}

    \caption{A semi-Markovian causal diagram}
    \label{fig:query-generation}
\end{figure}

As a simple example of query generation, consider the problem of generating all queries that either involve the causal effect on $Y$, i.e. $P(y \mid do(\ldots))$ or involve manipulating $x$, i.e. $P(\ldots \mid do(x))$), with $M=(\text{figure \ref{fig:query-generation}})$ and $I=P(v)$. Two such queries are identifiable: $Q_1 = P(y \mid do(x))$, and $Q_2=P(z \mid do(x))$.

Query generation can be combined with the other causal inference tasks. For example, starting with a known joint probability function, causal discovery can enumerate Markov compatible models, with query generation to enumerate identifiable queries for each Markov compatible model.

\section{Restricted causal inference relation}

The causal inference relation can be useful as a conceptual framework, but it is not a practical way of analyzing causal inference problems unless the domains of $M$, $I$ and $Q$ are appropriately restricted; note that if $M$ can include arbitrary model assumptions, then conducting causal inference may require invoking arbitrary mathematical theorems.

Several previously studied problems can be cleanly expressed as special cases of finding instances of the causal inference relation. In particular, identification has several subproblems that permit complete algorithms, in the sense that if it is possible to identify $Q$ from $M$ and $I$, then the algorithm is guaranteed to find an appropriate $F$. Let $G$ be a Markovian or semi-Markovian causal diagram, $P(v)$ be the joint observational probability function, and $W$, $X$, $Y$, and $Z$ each be subsets of $V$:

\begin{itemize}
    \item
        Causal effect identification (ID) \autocite{huang2006-identifiability, shpitser2006-identification}: $M = G$, $I = P(v)$, $Q = P(y \mid do(x))$

    \item
        Conditional causal effect identification (IDC) \autocite{shpitser2006-conditional}: $M = G$, $I = P(v) $, $Q = P(y \mid w, do(x))$

    \item
        Causal effect identification via surrogate experiments (zID) \autocite{bareinboim2012-surrogate}: $M = G$, $I = P(v \mid do(z')), \forall Z' \subseteq Z$, $Q = P(y \mid do(x))$

\end{itemize}

A zIDC algorithm, combining the capabilities of IDC and zID, would correspond to $M=G$, $I=P(v \mid do(z')), \forall Z' \subseteq Z$, $Q=P(y \mid w, do(x))$. Finding a complete algorithm for zIDC appears to be an open problem.

Causal discovery can be performed with Inductive Causation (IC) \autocite{pearl1991}. Given a probability distribution $P$ and assuming faithfulness, IC outputs a \emph{pattern}, which denotes an equivalence class of causal diagrams. If the underlying model is known to be Markovian, then IC is also complete, in that the resulting pattern will correspond to the complete set of causal diagrams that are Markov compatible with $P$. Otherwise, IC will produce a pattern that includes many, but not all, compatible semi-Markovian models.

\begin{figure}[h]
    \centering
    \includegraphics{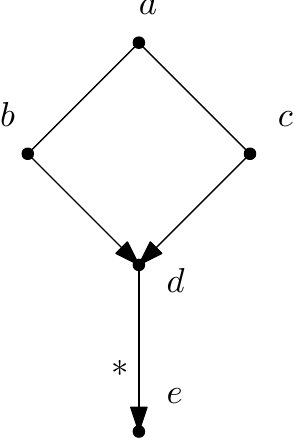}

    \caption{A marked pattern \autocite{pearl2009}. Marked edges, e.g. $d \rightarrow e$, signify a directed edge in the underlying model. Directed edges, e.g. $b \rightarrow d$, represent either $b \rightarrow d$ or a latent common cause of both $b$ and $d$. Undirected edges, e.g. $a \text{ --- } b$, represent either $a \leftarrow b$, $a \rightarrow b$, or a latent common cause.}
    \label{fig:pattern}
\end{figure}

Inductive Causation leaves the details of some its steps unspecified. In particular, IC requires searching for a set $S_{ab}$ such that $(a \ci b \mid S_{ab})$ for every pair of variables $a$ and $b$ in $V$, but does not specify how such sets should be found. The PC algorithm \autocite{spirtes1991-algorithm} is a refinement of IC that runs in polynomial time on fixed-degree graphs. The combination of IC-based algorithms and identification algorithms permit finding instances of the causal inference relation that correspond to known $I$ and $Q$. For example, the combination of PC and ID would permit finding instances of the causal inference relation that correspond to $I=P(v), Q=P(y \mid do(x))$

Problems related to research design have been discussed in the structural causal model literature; for example, Pearl notes that the front-door and back-door criteria permit an analyst a degree of freedom in selecting which set of covariates to adjust for when calculating causal effect \autocite{pearl2009}. However, the more general problem of finding instances of the causal inference relation corresponding to given $M$ and $Q$ does not appear to have an existing, standard formulation. Similarly, query generation is implicitly considered in the analysis of identification, but does not appear to have been formulated as a problem in its own right.

All of these problems can be unified as special cases of finding instances of the causal inference relation, with the following domains for $M$, $I$ and $Q$:

\begin{itemize}

    \item
        $M$: Markovian and semi-Markovian causal diagrams over $V$. Causal diagrams can be represented as $G=(V, E, C)$ where $(V, E)$ forms a directed acyclic graph and $C$ is a confounding family of $V$, corresponding to the dashed edges that represent latent confounding. A Sperner family is a collection of subsets of a given set, such that none of the subsets contain any of the others \autocite{weisstein-sperner}. This paper defines a confounding family of $V$ to be a Sperner family, with the further requirement that none of the subsets of $V$ are singleton.

    \item
        $I$: Information sets that can be represented as $P(w \mid do(z')), \forall Z' \subseteq Z$, for a given $Z \subseteq W$ and $W \subseteq V$. Note that this is simply $P(v)$ when $W=V$ and $Z=\emptyset$. This representation has a natural partial order: an information set $I_1$ is said to be contained in another information set $I_2$ if $W_1 \subseteq W_2$ and $Z_1 \subseteq Z_2$.
        
    \item
        $Q$: Queries that can be represented as $P(y \mid w, do(x))$ for a given $w \subseteq V, x \subseteq V, y \subseteq V$.

\end{itemize}

With respect to these domains, the following are suggested as canonical formulations of causal inference problems:

\begin{itemize}
        
    \item
        Identification (zIDC): Given $M, I, Q$, find one instance of the causal inference relation $\langle M, I, Q, F \rangle$.

    \item
        Causal discovery: Given $I, Q$, enumerate $\langle M_i, I, Q, F_i \rangle$ for distinct $M_i$, i.e. for any two enumerated instances of the causal inference relation, $\langle M_1, I, Q, F_1 \rangle$ and $\langle M_2, I, Q, F_2 \rangle$, $M_1 \neq M_2$

    \item
        Research design: Given $M, Q$, enumerate $\langle M, I_i, Q, F_i \rangle$, for distinct, minimal $I_i$, i.e. for any enumerated instance of the causal inference relation, $\langle M, I_i, Q, F_i \rangle$, there does not exist another instance $\langle M, I_j, Q, F_j \rangle$ such that $I_i$ is contained in $I_j$
        
    \item
        Query generation: Given $M, I$, enumerate $\langle M, I, Q_i, F_i \rangle$ for distinct $Q_i$

\end{itemize}

The combination of the axioms of probability theory and the inference rules of Pearl's causal calculus \autocite{pearl1995} are known to be complete for the ID, IDC and zID problems \autocite{huang2006-identifiability, shpitser2006-conditional, bareinboim2012-surrogate}. This paper conjectures that they are complete for all of the problems above as well. Furthermore, if a complete zIDC algorithm exists, it would constitute a complete --- albeit intractable, for larger $\vert V \vert$ --- solution for all of these problems. Causal discovery, research design and query generation problems can be reduced to identification problems by instantiating all possible Markov compatible causal diagrams, information sets, or queries, respectively, and running an identification algorithm for each instantiation.

\section{Other domains for the causal inference relation}

Other problems in causal inference can be represented in the causal inference relation framework by modifying the domain of the relation appropriately. In particular, the problems of identification of counterfactuals and recovery from selection bias require only minor extensions to $Q$ and $I$, respectively.

Identification of counterfactuals can be represented by extending the domain of possible queries. Let $G$ be a causal diagram, $\gamma$ and $\delta$ be conjunctions of counterfactual events, e.g. $Y_x, Z_w$, in the potential outcomes notation, and $P_*$ be the set of all experiments, i.e. $P(v \mid do(z')), \forall Z' \subseteq V$. With respect to these domains, the following problems are known to have complete algorithms \autocite{shpitser2008}:

\begin{itemize}

    \item
        Counterfactual identification (ID*): $M = G$, $I=P_*$, $Q=P(\gamma)$

    \item
        Conditional counterfactual identification (IDC*): $M=G$, $I=P_*$, $Q=P(\gamma \mid \delta)$

\end{itemize}

Selection bias can be represented by extending information sets to include ``s-biased'' data \autocite{bareinboim2012-controlling}, i.e. $P(v \mid S=1)$, where S represents a binary indicator of entry into the data pool. For example \autocite{bareinboim2014-thesis}, in studying the effect of a training program on earnings, subjects achieving higher incomes may tend to report their earnings more frequently than those who earn less. Recovery from selection bias is the problem of answering queries about the general population, despite the data being collected under selection bias. This data may be accompanied by unbiased data, $P(t)$, over some subset $T \subset V$. Bareinboim \autocite{bareinboim2014-thesis} outlines several problems related to selection bias:

\begin{itemize}

    \item
        Selection without external data: $M=G$, $I=P(v \mid S=1)$, $Q=P(y \mid x)$

    \item
        Selection with external data: $M=G$, $I=P(v \mid S=1), P(t)$, $Q=P(y \mid x)$
    
    \item
        Selection in causal inferences: $M=G$, $I=P(v \mid S=1), P(t)$, $Q=P(y \mid do(x))$

\end{itemize}

Complete identification criteria exist for selection without external data. Sufficient criteria and a valid algorithm for computing selection with external data and selection in causal inferences exist, but are not known to be complete. In particular, identification in the presence of both selection bias and latent confounding (i.e. in semi-Markovian models) is particularly difficult \autocite{bareinboim2014-thesis}.

Note that there are many causal inference problems that are not represented in this paper's formulation of the causal inference relation. There is no notion of providing bounds for query, when an exact result cannot be computed. Causal diagrams are the only form of model assumptions, which excludes parametric assumptions and nonrecursive (i.e. cyclic) systems. And the problem of external validity, i.e. generalizing results to a different environment from which the original data was collected, is not considered. In principle, the relation could be modified to represent these problems, but this would add considerable complexity. Introducing problems into the framework requires careful selection of the domains of $M$, $I$, $Q$ and $F$ to represent the problem of interest, while still permitting a small set of complete inference rules.

\section{Causal programming}

The causal inference relation casts problems in causal inference as the problem of finding instances of a logical relation. Causal programming is proposed as a further generalization of causal inference as an optimization problem. The problem is to find optimal instances of the causal inference relation with respect to a cost function:
\begin{align*}
    & \text{minimize} & g(M, I, Q) \\
    & \text{subject to} & \exists F : \langle M, I, Q, F \rangle \\
    & \text{and} & M \in M^*, I \in I^*, Q \in Q^*
\end{align*}

Where $g$ is a cost function, $\exists F : \langle M, I, Q, F \rangle $ is the statement that there exists a formula such that $\langle M, I, Q, F \rangle$ constitutes an instance of the causal inference relation, and $M^*$, $I^*$, and $Q^*$ are the given domains for models, information sets and queries under consideration.

A natural problem to consider in this framework is the problem of optimal research design. For example, consider a scenario where an analyst wishes to calculate $P(y \mid do(x))$ with respect to the causal diagram in figure \ref{fig:research-design}. Since $M$ and $Q$ are given, the only degree of freedom is in the domain of information sets; $M^*$ is just a single causal diagram, i.e. $M^* = (\text{figure \ref{fig:research-design}})$ and $Q^*$ is just a single query, i.e. $Q^* = P(y \mid do(x))$.

Let the domain of information sets be all joint probability functions over subsets of $V$, i.e. $I^* = P(w), \forall W \subseteq V$, and the cost function, $g(I)$, be a linear cost function where including each $w_i$ in the joint probability function costs $i$, e.g. the cost of $P(v) = P(x, y, w_1, w_2, \ldots w_6)$ is $21$. In this example, the solution would be the instance of the causal inference relation: $\langle M, I, Q, F \rangle$, where $M$ and $Q$ are as given, $I = P(x, y, w_3, w_4)$, and $F = \sum_{x_3, x_4} P(y \mid x, w_3, w_4) P(w_3, w_4)$, with a cost of $7$.

As a function of $I$, $g$ can be interpreted as the cost of performing observational and/or experimental research, with the optimal instance of the causal inference relation representing the least expensive way to answer the original query. As a function of $M$, $g$ can be interpreted as the complexity of a model, with the optimal solution representing the simplest set of additional assumptions that permit answering the original query --- a formalization of Occam's razor. Finally, as a function of $Q$, $g$ can be interpreted as the (inverse, when minimizing $g$) value of being able to identify a particular query, which can be combined with other causal inference tasks. For example, given a causal model, but not an information set or query, finding an optimal instance of the causal inference relation would represent the finding the most valuable, identifiable query, and the information set required to compute it.

\section{Discussion}

Consider the steps involved in an idealized, simplified scientific method: Observe. Hypothesize. Predict. Experiment. (Repeat.) This corresponds well to tasks associated with the causal inference relation. Observation corresponds to obtaining an information set that includes observational probabilities. Hypothesizing corresponds to causal discovery of compatible models. Prediction requires generating an identifiable query. Finally, experimentation corresponds to obtaining interventional probabilities that confirm or deny the prediction. This process can be repeated with the interventional probabilities included in the information set.

This formalization relies on two conceptual limitations: that the variables under consideration all belong to a fixed set of endogenous variables, and that the analyst has access to population probability functions over these variables. However, within these restrictions, the causal inference relation and causal programming frameworks are powerful conceptual tools, providing means to represent many existing tasks in causal inference, and define novel ones, such as the (optimal) research design problems. The challenge remains to select domains for the causal inference relation that are expressive enough to unify many problems of interest while still permitting complete inference rules and tractable algorithms. Implementing a causal programming solver would constitute an important step towards building intelligent systems to automate the process of scientific discovery.

\section*{Acknowledgements}

Thanks to Alex Eftimiades and Brendan Good for reading early drafts of this paper.

\printbibliography

\end{document}